\documentclass[aps,pra,reprint,amsmath,amssymb,bibnotes,twocolumn]{revtex4-2}
\usepackage{xcolor,color,graphicx,float,hyperref,amsmath}

\usepackage{amsmath}
\usepackage{graphicx}
\def\ds{\displaystyle}

\begin{document}
\title{Edge burst effect and scale-free localization}
\author{G. Sen, C. Yuce}
\affiliation{Department of Physics, Eskişehir Technical University, Eskişehir 26555, Türkiye}

\begin{abstract}
The non-Hermitian edge burst is a phenomenon observed in non-Hermitian quantum dynamics, characterized by a significant accumulation of loss at the boundaries of a system. We present an example of the edge burst effect in a lossy lattice with bipolar non-Hermitian skin effect (NHSE). By introducing a weak coupling between two such non-Hermitian lattices, we demonstrate that the system exhibits bipolar scale-free localization. Through an analysis of local decay distributions and their sensitivity to system parameters, we confirm the occurrence of the edge burst effect in the system displaying scale-free localization. Our findings support the notion that the skin effect is not a necessary condition for the edge burst effect.
\end{abstract}

\maketitle

\section{Introduction}

Non-Hermitian systems in one dimension exhibit two unique types of localizes states: skin and scale-free localized (SFL) states. Skin states, localized at the edges, are characterized by size-independent localization lengths and are induced by NHSE that is linked to the spectral winding number, a topological invariant unique to non-Hermitian systems \cite{sw1,sw2,sw3,slager,ref2ek1,ref2ek2,ref2ek3,ref2ek4,ref2ek5,ref2ek6}. Conversely, SFL states are localized states whose localization lengths scale proportionally with the system size. SFL states appear in a non-Hermitian ladder lattice or even in a Hermitian lattice with a single non-Hermitian defect  without necessarily relying on the non-Hermitian topology \cite{sflA,sfl0,sfl1,sfl2,sfl3,sfl4,sfl5,sfl8,sfl9,sfl10,sflcem,sflwang}. SFL states has recently been experimentally realized in circuit lattices with non-Hermitian defects \cite{sfl8}.\\
Non-Hermitian systems also demonstrate two intriguing dynamical effects: non-Hermitian funneling effect \cite{scifun,yucefunnel,scifun2,scifun3} and edge burst effect \cite{in1,in2,in3,in3cem,in4,offbea,offbea2,offbea3}. The non-Hermitian funneling effect, caused by the NHSE, is observed in the context of light propagation in photonic lattices, where light is always transported to a focal point regardless of its initial form. The recent proposal of the non-Hermitian edge burst \cite{in1,in2} has sparked interest among researchers. It refers to a significant amount of loss occurring at the system boundary in a lossy lattice, even when the initial single-site excitation is far from the edge. Understanding and characterizing this behavior is crucial for both fundamental research and practical applications. Initially, the edge burst effect was linked to topological edge states \cite{in1}, but a recent paper has challenged this connection by proposing that this effect appears when NHSE and imaginary gap closing occur in the system \cite{in2}. However, we considered a specific non-Hermitian lattice with nonuniform loss rates and found that while these conditions can give rise to the edge burst, they are not necessary and sufficient for the emergence of the edge burst effect \cite{in3}. This effect has recently been studied in a lattice with asymmetrical long-range couplings, exhibiting bipolar NHSE and it was shown that asymmetrical coupling leads to strong edge burst effect \cite{in3cem}. Peng Xue's group has experimentally confirmed the existence of the edge burst effect through single-photon discrete-time quantum walks \cite{in4}, and this effect has also been observed in one-dimensional photonic quantum walks \cite{offbea}.\\
To date, it remains unclear whether a system with SFL states can exhibit an edge burst. Our paper aims to investigate this issue and demonstrate an example of the edge burst effect in a system exhibiting scale-free localization. We consider a lossy lattice without asymmetrical coupling that exhibits bipolar NHSE. Introducing coupling between two such non-Hermitian lattices destroys the bipolar NHSE and leads to the emergence of bipolar scale-free localization. Each sublattice hosts SFL modes localized at both edges, unlike the coupled Hatano-Nelson model \cite{sfl0}, where SFL modes are localized only at one edge of each sublattice. We show that the edge burst effect occurs in this system by analyzing local decay distribution, and its sensitivity to system parameters. This supports the argument in the literature that NHSE is not necessary for the occurrence of edge bursts. This paper contributes to the knowledge on the non-Hermitian edge burst effect.

 \section{Bipolar scale-free localization}

NHSE can occur without non-reciprocal (asymmetrical) couplings \cite{in2}. In one dimension, this can happen when the system contains sub-systems with equal and opposite winding number. Here, we propose a one-dimensional lossy tight-binding lattice that can exhibit either skin or scale-free localization, depending on the system parameters. When gain/loss is the only source of non-Hermiticity, skin localization may occur if there is a phase shift between nearest-neighbor and long-range couplings. In contrast, scale-free localization can be induced by coupling two identical lattices that exhibit skin localization. Fig. 1 depicts our model with two real-valued parameters, $\gamma$ and $\Delta$, which control skin and scale-free localization, respectively. The system consists of two weakly coupled Hermitian sublattices $A$ and $B$ with a coupling parameter $\Delta$, $0<\Delta<<1$, and the couplings between nearest and next-to-nearest neighboring sites in each sublattice are set to $1$ and ${{\mp}i\gamma}$, respectively. We also introduce losses with a loss rate $V$ to the odd-numbered sites in each sublattice. When the sublattices are uncoupled ($\Delta=0$), each sublattice separately displays bipolar NHSE \cite{in3cem}. In contrast, coupling the sublattices destroys the bipolar NHSE and leads to bipolar scale-free localization. This minimal model is the basis for our investigation on the non-Hermitian edge burst phenomena in a system exhibiting scale-free localization. The model equations for the field amplitudes $\ds{\psi_j^A}$ and $\ds{\psi_j^B}$ at the sublattices $A$ and $B$ for site $j=1,2,...,N$ are given by 
	\begin{eqnarray}\label{1eqtw461}
		\psi_{j+1}^A + \psi_{j-1}^A+\Delta  \psi_{j}^B  + i  \gamma( \psi_{j+2}^A  -  \psi_{j-2}^A   ) - i
		V_{j}  \psi_{j}^A  =i \frac{d \psi_{j}^A  }{d t} \nonumber\\
		\psi_{j+1}^B + \psi_{j-1}^B+\Delta  \psi_{j}^A - i \gamma (\psi_{j+2}^B  -  \psi_{j-2}^B   ) - i
		V_{j}  \psi_{j}^B   = i \frac{d \psi_{j}^B  }{d t}
	\end{eqnarray}
	where the loss rates, $V_{j}>0$, are nonzero only at odd-numbered sites, i. e., $V_j=V $ when $j$ is odd and $V_j=0 $ when $j$ is even. We solve these equations by imposing the open boundary conditions (OBC), $\psi_{N+2}^{A,B}=\psi_{N+1}^{A,B}=\psi_{0}^{A,B}=\psi_{-1}^{A,B}=0$. To study the spectrum of the system, we write $\psi_{j}^{A,B}(t)=e^{-iEt}~\Psi_{j}^{A,B}$, where $E$ and $\Psi_{j}^{A,B}$ are eigenvalues and eigenstates, respectively.\\
The type of localization, namely skin and scale-free localization, are controlled by the parameters $\gamma$ and $\Delta$, respectively. When $\Delta=0$, where the system is decoupled into two distinct non-Hermitian lattices, each lattice exhibits bipolar NHSE \cite{in3cem}. This means that skin states can accumulate at both boundaries rather than being confined to just one. This bipolar NHSE can be predicted by the spectral winding number \cite{bipolar2}: ${	\nu = \int_{-\pi}^{ \pi} \frac{dk}{2\pi{i}} \frac{d }{dk} ln [ E_{\mp} (k)-E_0 ]}$ for a complex reference energy $E_0$, where the spectrum under periodic boundary conditions (PBC) is given by ${E_{\mp} (k) = - 2\gamma ~\sin(k)-i \frac{V}{2} \mp \sqrt{2 - \frac{V^2}{4} + 2 \cos(k)} } $ with $-\pi{\leq}k<\pi$. In the complex energy plane, this expression for $E_{\mp} (k)$ determines an upper and lower loop that intersect along a line (known as $\it{Bloch ~ line}$). For example, at ${V=2\sqrt{2}}$, this expression simplifies to ${E_{\mp} (k) = - 2\gamma ~\sin(k)-\sqrt{2}i \mp\sqrt{2 \cos(k) }} $, which traces a straight line in the complex energy plane for ${-\pi/2{\leq}k\leq\pi/2}$, and forms upper/lower loops outside this range. Note that the winding number is equal to $\mp1$ if $E_0$ lies within the upper or lower loop. The opposite signs of the winding numbers of the loops indicate the presence of a bipolar NHSE in the system. The corresponding OBC modes with eigenvalues inside the PBC loops are identified as skin states. The remaining OBC states with the eigenvalues lying on the Bloch line are extended. Note that the length of the Bloch line decreases with increasing $V$, leading to more skin states and less extended states. At the critical loss rate ${V=4}$, the Bloch line becomes a Bloch point, as the upper and lower loops touch at a single point. This is evident from the expression ${E_{\mp} (k) = - 2\gamma ~\sin(k) -2i\mp2i \sin(k/2) } $, from which it follows that ${E_{+} (k=0)=E_{-} (k=0)  }$, indicating the Bloch point ${k=0}$. When $V>4$, the Bloch point disappears as the PBC loops don't touch each other (an imaginary line gap arises), and the system has only skin modes. \\
\begin{figure}[t]
	\includegraphics[width=4.5cm]{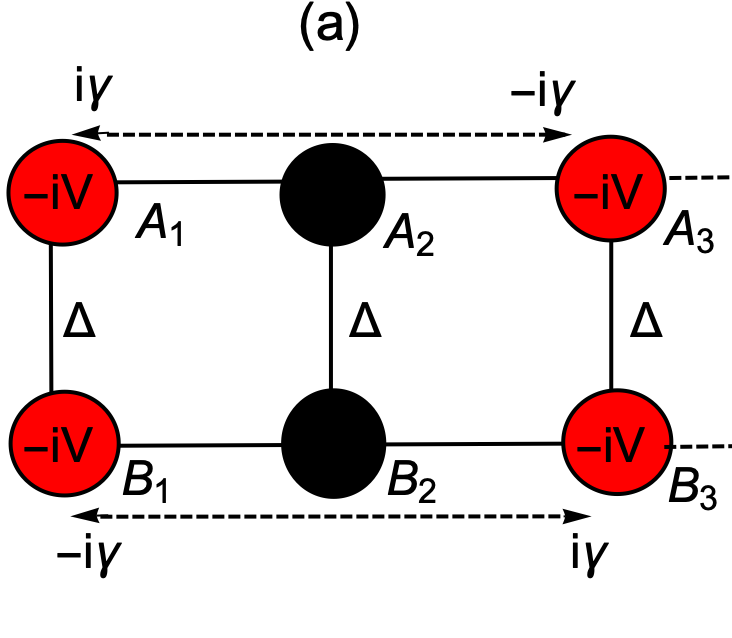}
        \includegraphics[width=4.25cm]{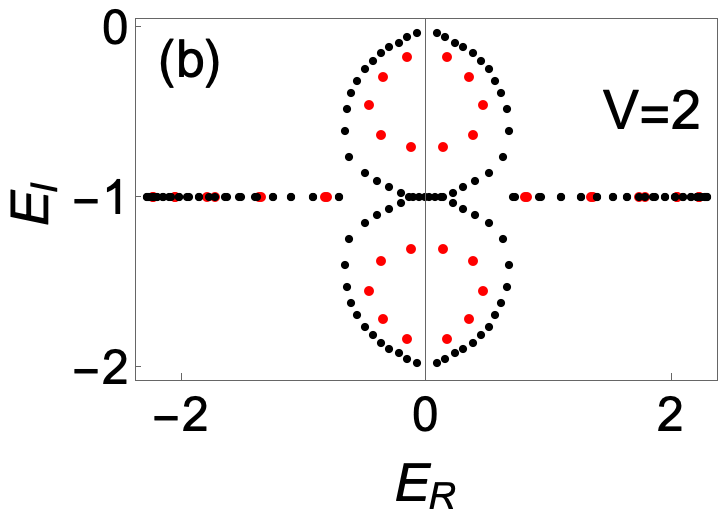}
	\includegraphics[width=4.25cm]{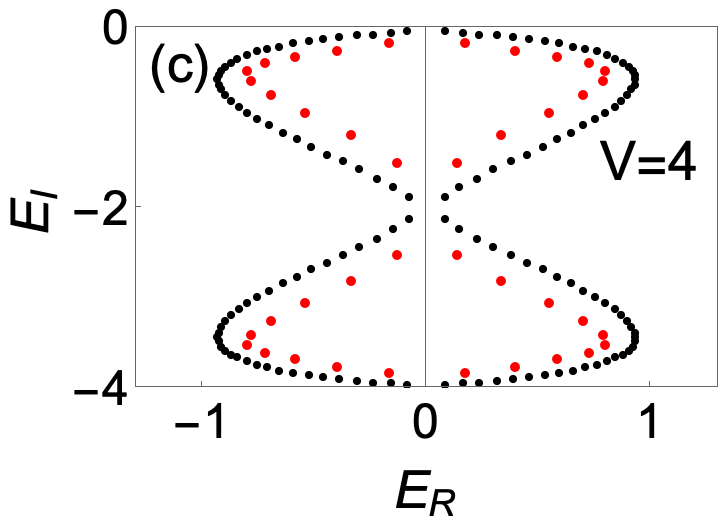}
	\caption{(a) The sublattices $A$ and $B$ are coupled with the coupling constant $\Delta$. The lattice points in red are lossy with the loss rate $V$, while the one in black are lossles. The couplings between adjacent sites are equal to $1$,  while the couplings between next-to-nearest neighboring sites are purely imaginary. The OBC spectra are presented in panel (b) for $V=2$ and in panel (c) for $V=4$. These spectra depend on the system size, with results shown for $N=20$ (red) and $N=70$ (black). The OBC spectral loops gradually expand with system size and approach the corresponding PBC loops in the limit of $N\rightarrow\infty$. The parameters used are $\gamma=0.5$, $\Delta=0.05$.}
\end{figure}
Coupling the sublattices ($\Delta\neq0$) causes the bipolar NHSE—and thus the skin modes under OBC—to disappear. Instead, bipolar scale-free localization occurs under OBC, hence each sublattice exhibits eigenstates with size dependent localization lengths. A specific SFL state is localized at one edge in sublattice $A$ and at the opposite edge in sublattice $B$ (asymmetric localization between the sublattices). Furthermore, in each sublattice, half of the SFL states are localized at the left edge and the other half at the right edge (bipolar localization in each sublattice. Note that the analogy between NHSE and bipolar NHSE can be extended to scale-free localization and bipolar scale-free localization). As a result, the OBC spectrum of the coupled system becomes highly sensitive to the lattice size. Fig. 1 (b,c) illustrates this, where we see substantial differences in the OBC spectra when ${N=20}$ (red) and ${N=70}$ (black). The OBC spectrum consists of upper and lower loops that gradually expand with system size and approach the corresponding PBC loops in the limit of $N\rightarrow \infty$, where the upper loop touches the $E_R$ axis (a zero-energy OBC eigenstate exists, and the imaginary gap closes). Note that Bloch line appears for the system in panel (b), indicating that the OBC systems exhibit a coexistence of extended and bipolar SFL states. \\
Let us illustrate the SFL states by plotting their densities when $N=20$ and $V=4$. A specific SFL state is distributed between the two sublattices in such a way that it appears at opposite edges in each sublattice with $\sum_j |\psi_j^A|^2+|\psi_j^B|^2=1$. On the other hand, bipolar scale-free localization implies that each sublattice hosts bipolar SFL states, with half of these SFL states localized at the left edge and the other half at the right edge. Fig. 2(a) shows the densities of the SFL states localized at the left edge of sublattice $A$ and the right edge of sublattice $B$, while the inset plots for the other half of the SFL states. Note that all the states in the system are SFL states at $V=4$. If we increase the lattice size while keeping all other parameters fixed, the SFL states extend further toward the opposite edge in each sublattice. To see that their localization lengths change proportionally with system size, Fig. 2(b) plots ${\frac{<j>}{N}=\frac{\sum_j  j~(|\psi_j^A|^2+|\psi_j^B|^2) }{N}}$ for the SFL state with the lowest imaginary eigenvalue as a function of lattice size at $V=2$ and $V=4$. As can be seen, $\frac{<j>}{N}$ remain almost the same regardless of the system size and change little with the potentials for $V=2$ (in blue) and $V=4$ (in red). It is well known that spectral localization in non-Hermitian systems does not always lead to the suppression of wavepacket spreading (dynamical localization) \cite{yucefunnel}. In our system, an initially localized wave packet tends to move asymmetrically toward one edge of the sublattice $A$ and the opposite edge of the sublattice $B$, depending on the sign of ${\gamma}$. However, when ${\gamma = 0}$, this asymmetry disappears, and the wave packet spreads to both edges in both sublattices. This asymmetric motion leads to distinct decay characteristics for the sublattices. Below, we show that the non-Hermitian edge burst effect appears in our system, exhibiting scale-free localization. This supports the argument in the literature that the existence of skin modes due to NHSE are not necessary for the occurrence of edge burst.
 \begin{figure}[t]
\includegraphics[width=4.15cm]{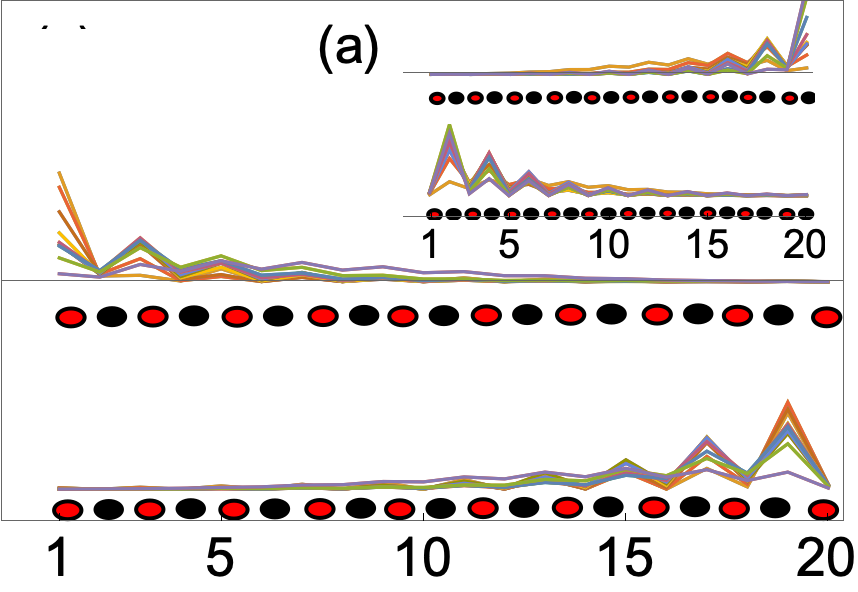}
\includegraphics[width=4.4cm]{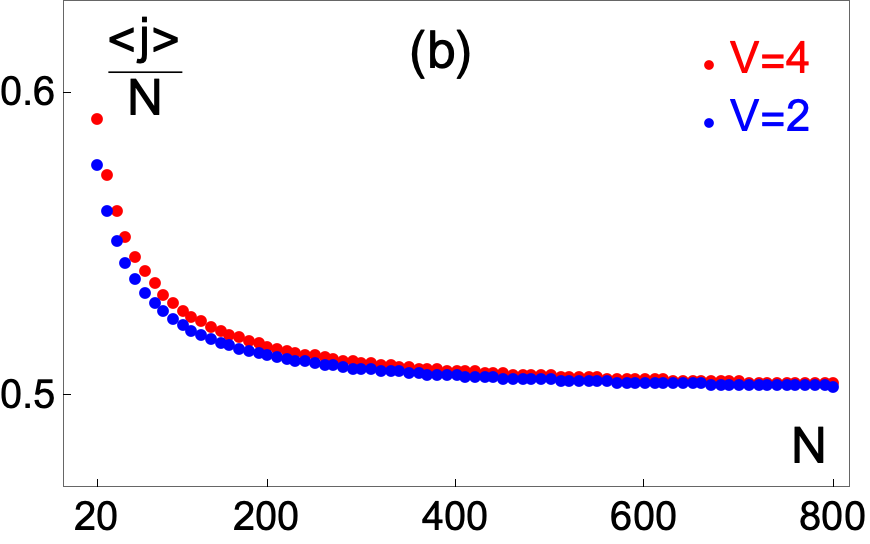}
	\caption{(a) Density profiles of SFL modes as a function of site index $j$ are shown. The left-localized SFL modes appear on sublattice $A$ (upper lattice), while the right-localized modes are on sublattice $B$ (lower lattice). In the inset, this configuration is reversed, reflecting the bipolar nature of SFL states within each sublattice. Each sublattice exhibits opposite-edge localization for a given SFL state with the normalization condition $\sum_j |\psi_j^A|^2 + |\psi_j^B|^2 = 1$. Lossy and lossless sites are indicated by red and black circles, respectively. (b) $\frac{<j>}{N}=\frac{\sum_j  j~(|\psi_j^A|^2+|\psi_j^B|^2) }{N}$ as a function  of $N$ for the SFL state with the lowest imaginary eigenvalue at $V=2$ (in blue) and $V=4$ (in red).}
\end{figure} 
\begin{figure}[t]
	\includegraphics[width=4.25cm]{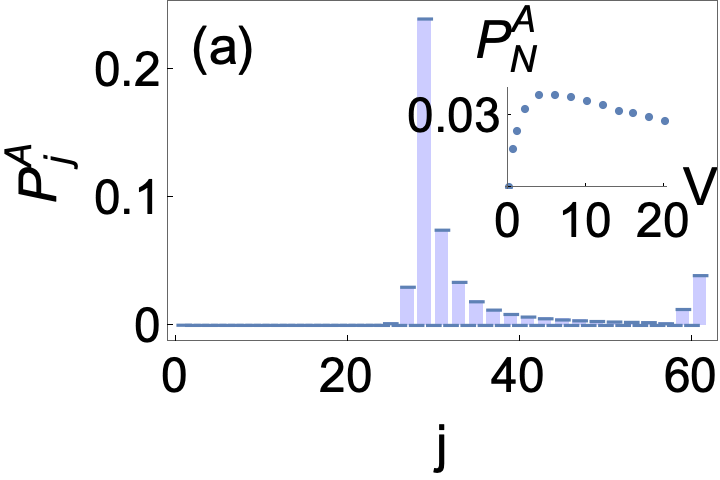}
	\includegraphics[width=4.25cm]{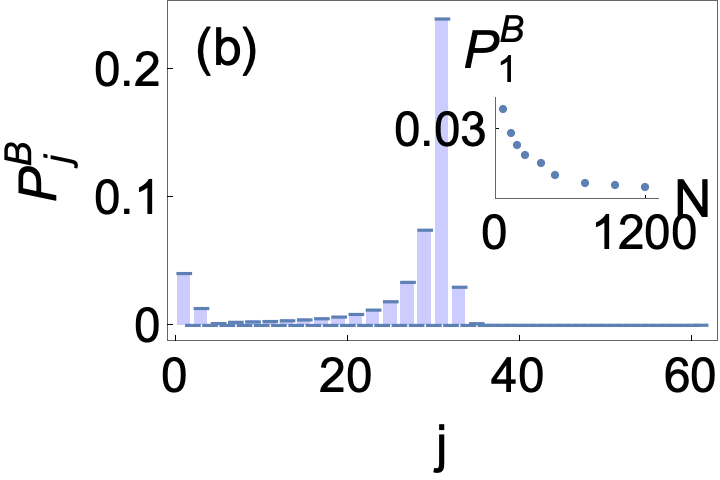}
 	\includegraphics[width=4.25cm]{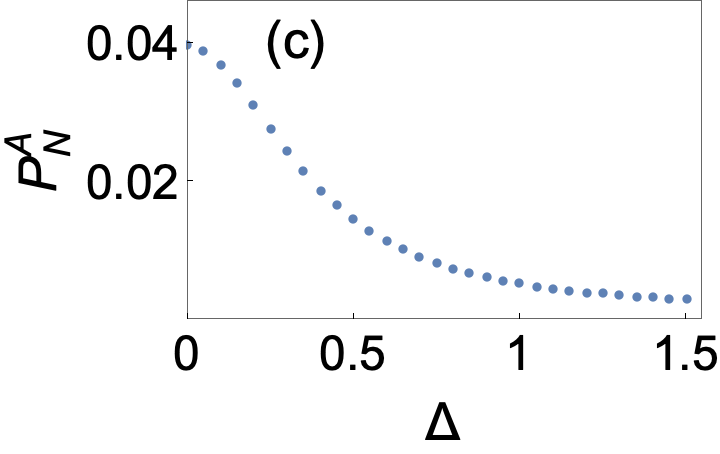}
         \includegraphics[width=4.25cm]{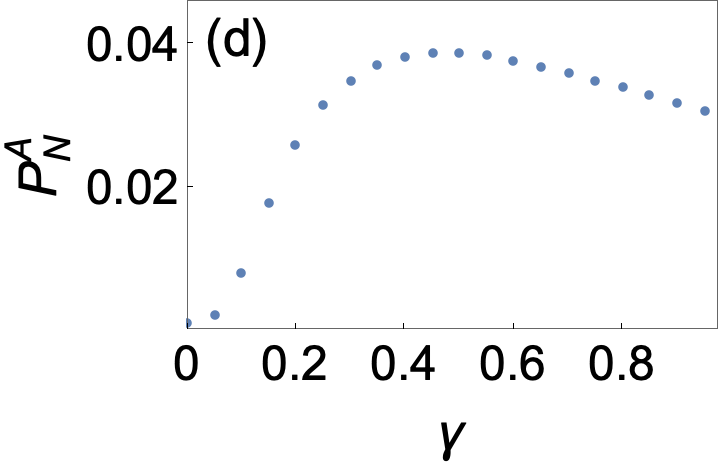}
		\caption{(a) and (b) illustrates the edge burst at $\gamma=0.5$, $\Delta=0.05$ and $V=4$ for sublattice $A$ and $B$, respectively. The total loss from the edge can be enhanced by varying the loss rate $V$, as can be seen from the inset of (a), plotting $P_N^{A}$ as a function of $V$. In the inset of (b), $P_1^{B}$ as a function of lattice size when $S=\frac{N-1}{2}$. (c) $P_N^{A}$ as a function of $\Delta$ at $\gamma=0.5$. (d) $P_N^{A}$ as a function of $\gamma$ at $\Delta=0.05$. The common parameters are $N=61$, $V=4$, $\psi^A_j(t=0)=\psi^B_j(t=0)=\frac{\delta_{j,30}}{\sqrt{2}}$.}
\end{figure}

\section{Non-Hermitian Edge Burst}

The edge burst effect refers to a significant loss at one boundary of the system \cite{in1,in2,in3,in3cem}. In our model,  SFL states show symmetric localization at opposite sublattice edges, and each sublattice hosts them, half localized at the right edge and half at the left; hence, one may expect a symmetrical edge burst effect in each sublattice. However, a significant portion of the loss occurs at the opposite edges of the sublattices. The funneling effect causes the initial wave to propagate toward the opposite edges of the sublattices, implying an asymmetric dissipation in each sublattice.\\
Suppose that the system with an odd-number lattice size is initially excited at a single lossless site in each sublattice, described by the initial conditions ${\psi^A_j(t=0)=\psi^B_j(t=0)=\frac{\delta_{j,S}}{\sqrt{2}}}$, where the starting site $S$ is an even number. Both sublattices are lossy, causing the wave to completely leak out of the system as $t\rightarrow \infty$. The total loss from each site in the sublattices $A$ and $B$ are determined by the local decay probabilities. From Eq. (1), we get $ { I^{A,B}=-\int_0^t~\sum_j 2 V_j ~ |\psi_j^{A,B}|^2  ~dt}$, where ${I^{A,B}=\sum_j |\psi_j^{A,B}|^2}$ is the total power for each sublattice \cite{in3cem}. Therefore, we obtain the local decay probabilities for sublattice $A$ and $B$
\begin{eqnarray}\label{ket5d6}
 P_j^{A,B} = 2~V_j\int_0^{\infty}|\psi_{j}^{A,B}|^{2} ~dt 
\end{eqnarray}
where the total dissipated power is equal to the initial
power $\sum_j P_j^{A}+P_j^{B}=1 $. Note that the local decay probabilities are non-vanishing only when $j$ is an odd number since losses are not introduced at the even-numbered sites. When ${\gamma>0}$, a large portion of the decay occurs at right and left edges of the sublattice $A$ and $B$, respectively (this is reversed when $\gamma<0$). Therefore, we define the following conditions as evidence for the presence of the edge burst in each sublattice: ${P_N^{A}/P_{min}^{A}>>1}$ and ${P_1^{B}/P_{min}^{B}>>1}$, where $P_{min}^{A} = min  \{P_{S+1}^{A},P_{S+3}^{A}, ..., P_N^{A}  \}$ and $P_{min}^{B} = min  \{P_1^{B},P_3^{B},..., P_{S-1}^{B}  \}$ represent the minimum of the local decay probabilities between the starting point $S$ and the right and left edge for sublattice $A$ and $B$, respectively \cite{in2}. In the following, we numerically solve Eq.(\ref{1eqtw461}) and verify that these conditions are satisfied, demonstrating the non-Hermitian edge burst effect in the system.\\
We illustrate the edge burst effect in the system with $V=4$, which exhibits only bipolar SFL states. It is worth noting that this effect also occurs for smaller values of $V$, where the system exhibits a coexistence of bipolar SFL and extended states. Figure 3(a,b) plot the local decay probability distributions when $N=61$ and $S=30$. We find $P_{N+1-j}^{A}{\approx}P_{j}^{B} $, where the slight difference is due to odd number of $N$. As can be seen, $P_j^{A,B}$ takes its maximum value near the initially populated site and rapidly decreases asymmetrically in the bulk towards the edges. Remarkably, local decay probabilities exhibit a sharp peak at the right edge of sublattice $A$ and the left edge of sublattice $B$, indicating an edge burst effect. Conversely, these probabilities are almost zero at the left edge of sublattice $A$ and the right edge of sublattice $B$. \\
The edge burst effect can be enhanced by varying the loss rate $V$ or changing the system size. For instance, we compute $P_N^{A}$ for the above illustration as a function of $V$, and plot it in the inset of Fig.3(a). As can be seen, it is zero at ${V=0}$ and increases dramatically with the loss rate $V$ until $V\approx5$, from which it starts to slowly decrease with it. This can be understood as follows. Increasing the loss rate $V$ enhances the funneling effect that directs the initial wave to the edge. On the other hand, a higher loss rate also leads to more losses near the starting lattice site. This competition causes $P_N^{A}$ to increase with $V$, followed by a decline as $V$ is increased further. Next, we compute the local decay probabilities for various system size to show that the lattice size also changes the strength of the edge burst effect. In the inset of Fig.3(b), we plot $P_1^{B}$ as a function of lattice size when ${S=\frac{N-1}{2}}$ up to ${N=1200}$, showing that edge leakage decreases with increasing lattice size. Since the initial site is farther from the edge, the wave packet spends more time and decays more before arriving at the edge.\\
To explore how edge losses depend on the parameters $\Delta$ and $\gamma$, we consider the sublattice $A$ and plot $P_N^{A}$ as a function of $\Delta$ (at $\gamma=0.5$) in panel (c) and as a function of $\gamma$ (at $\Delta=0.05$) in panel (d). As seen in Fig.3(c), $P_N^{A}$ reaches its maximum when the sublattices are uncoupled (when NHSE occurs). Introducing the coupling, $\Delta\neq0$, where scale-free localization occurs, $P_N^{A}$ does not sharply drop to zero, but instead decreases gradually to zero as $\Delta$ increases. We conclude that the edge burst effect is stronger in the presence of the skin localization, but continues to exist at a reduced level under scale-free localization. For large values of $\Delta$, non-Hermitian edge burst effect does not practically occur. This is because the OBC and PBC spectra converge as $\Delta$ increases, destroying scale-free localization (The localization lengths of OBC states become as large as our lattice size for large values of $\Delta$, hence the OBC states become effectively extended states). Finally, we stress that the edge burst effect does not occur in our system without the coupling of next-to-nearest neighboring sites. Fig.3(d) demonstrates how the parameter $\gamma$ changes the strength of the edge burst effect by computing $P_N^{A}$ as a function of $\gamma$. $P_N^{A}$ is zero when ${\gamma=0}$, indicating the absence of the edge burst effect in the absence of next-to-nearest neighboring coupling. Furthermore, $P_N^{A}$ increases with $\gamma$. However, this growth is not unbounded. It reaches a maximum near ${\gamma=0.45}$ (this number increases slightly with $N$). Beyond this point, $P_N^{A}$ decreases with $\gamma$.\\
A positive or negative value of the next-to-nearest neighbor coupling directs the initial wave packet toward the right or left edge, respectively, implying that the sign of $\gamma$ determines the direction of funneling effect and the edge from which a significant amount of losses occurs. In the illustration presented above, but with the sign of $\gamma$ reversed, ${P_j^{A}}$ and ${P_j^{B}}$ are exchanged. This is evident from the interchange of $\psi_j^A(t)$ and $\psi_j^B(t)$, as shown in Eq. (\ref{1eqtw461}). It is worth noting that if the sign of $\gamma$ is reversed only for sublattice $B$, significant losses occur at the right edge for both sublattices. This occurs because $\ds{\psi_j^A(t)=\psi_j^B(t)}$ according to Eq.(\ref{1eqtw461}), which in turn implies ${P_j^{A}=P_j^{B}}$. On the other hand, if it is zero only for sublattice $B$, then only the sublattice $A$ exhibits an edge burst effect (sublattice $B$ shows a symmetric decay distribution around the initial site $S$ without any sharp peak at the edge).

\section{Conclusion}

The edge burst is a novel phenomenon in non-Hermitian dynamics that has recently attracted a great deal of attention. This effect implies large losses at the boundaries of the systems. It was suggested that the edge burst arises from the interplay between two critical non-Hermitian phenomena: NHSE and imaginary gap closing. This paper studies the edge burst effect in the system without skin localizations. The system we consider displays bipolar scale-free localization, where SFL states are localized at both edges with size-dependent localization lengths. We show that the edge burst asymmetrically occurs in each sublattice. We think that the asymmetrical local loss distribution is due to the funneling effect, which can occur in the presence of scale-free localization. We discuss that the edge burst effect reduces with the system size. We also explore how the edge burst enhances with the system parameters.\\
Although scale-free localization does not prevent the edge burst from occurring, not all systems exhibiting scale-free localization displays this effect. For instance, a coupling impurity in an otherwise Hermitian lattice can induce scale-free localized states \cite{sfl2}. Adding uniform losses to each site does not change the spectra and the form of the eigenstates. However, the single coupling impurity does not cause the system to exhibit the edge burst effect. \\
We thank Dr. Hamidreza Ramezani for the valuable discussion. This study was supported by Scientific and Technological Research Council of Turkey (TUBITAK) under the grant number 123N121. The authors thank to TUBITAK for their supports.


\begin{thebibliography}{0}

\bibitem{sw1} Huitao Shen, Bo Zhen, and Liang Fu, \textquotedblleft{Topological Band Theory for Non-Hermitian Hamiltonians\textquotedblright}, Phys. Rev.	Lett. {\bf 120}, 146402 (2018).
\bibitem{sw2} Nobuyuki Okuma, Kohei Kawabata, Ken Shiozaki, and Masatoshi Sato, \textquotedblleft{Topological Origin of Non-Hermitian Skin Effects\textquotedblright}, Phys. Rev. Lett.  {\bf 124}, 086801(2020).
\bibitem{sw3} Kai Zhang, Zhesen Yang, and Chen Fang, \textquotedblleft{Correspondence between Winding Numbers and Skin Modes in Non-Hermitian Systems\textquotedblright}, Phys. Rev. Lett.  {\bf 125}, 126402 (2020).
\bibitem{slager} Dan S. Borgnia, Alex Jura Kruchkov, and Robert-Jan Slager,\textquotedblleft{Non-Hermitian Boundary Modes and Topology\textquotedblright}, Phys. Rev. Lett. {\bf 124}, 056802 (2020).
\bibitem{ref2ek1} Jinbing Hu, et. al., \textquotedblleft{Electric polarization and its quantization in one-dimensional non-Hermitian chains\textquotedblright}, Phys. Rev. B {\bf107}, L121101 (2023).
\bibitem{ref2ek2} Ye Xiong, \textquotedblleft{Why does bulk boundary correspondence fail in some non-hermitian topological models\textquotedblright}, J. Phys. Commun. 2 035043 (2018).
\bibitem{ref2ek3} Zongping Gong, \textquotedblleft{Topological Phases of Non-Hermitian Systems\textquotedblright}, Phys. Rev. X {\bf8}, 031079 (2018).
\bibitem{ref2ek4} Kohei Kawabata, Ken Shiozaki, Masahito Ueda, and Masatoshi Sato, \textquotedblleft{Symmetry and Topology in Non-Hermitian Physics\textquotedblright}, Phys. Rev. X {\bf9}, 041015 (2019).
\bibitem{ref2ek5} Shunyu Yao and Zhong Wang, \textquotedblleft{Edge States and Topological Invariants of Non-Hermitian Systems\textquotedblright}, Phys. Rev. Lett. {\bf121}, 086803 (2018).
\bibitem{ref2ek6} Tony E. Lee, \textquotedblleft{Anomalous Edge State in a Non-Hermitian Lattice\textquotedblright}, Phys. Rev. Lett. {\bf116}, 133903 (2016).
\bibitem{sflA} L. Li, C. H. Lee, S. Mu, and J. Gong, \textquotedblleft{Critical non-Hermitian skin effect\textquotedblright}, Nat. Commun. {\bf11}, 5491 (2020).
\bibitem{sfl0} Kazuki Yokomizo, Shuichi Murakami, \textquotedblleft{Scaling rule in critical non-Hermitian skin effect\textquotedblright}, Phys. Rev. B {\bf104}, 165117 (2021).
\bibitem{sfl1} Bo Li, He-Ran Wang, Fei Song, and Zhong Wang, \textquotedblleft{Scale-free localization and	PT symmetry breaking from local non-Hermiticity\textquotedblright}, Phys. Rev. B {\bf108}, L161409 (2023).
\bibitem{sfl2} Cui-Xian Guo, Xueliang Wang, Haiping Hu, and Shu Chen, \textquotedblleft{Accumulation of scale-free localized states induced by local non-Hermiticity\textquotedblright}, Phys. Rev. B {\bf107}, 134121  (2023).
\bibitem{sfl3} Linhu Li, Ching Hua Lee, Jiangbin Gong, \textquotedblleft{Impurity induced scale-free localization\textquotedblright}, Communications Physics {\bf4}, 42 (2021).
\bibitem{sfl4} Yongxu Fu, Yi Zhang, \textquotedblleft{Hybrid scale-free skin effect in non-Hermitian systems: A transfer matrix approach\textquotedblright}, Phys. Rev. B {\bf108}, 205423 (2023).
\bibitem{sfl5} Wei Li, Zhoujian Sun, Ze Yang, and Fuxiang Li, \textquotedblleft{Universal scalefree non-Hermitian skin effect near the Bloch point\textquotedblright}, Phys. Rev. B {\bf109}, 035119 (2024).	
\bibitem{sfl8} Xinrong Xie, Gan Liang, Fei Ma, Yulin Du, Yiwei Peng, Erping Li, Hongsheng Chen, Linhu Li, Fei Gao, Haoran Xue, \textquotedblleft{Observation of scale-free localized states induced by non-Hermitian defects\textquotedblright}, Phys. Rev. B {\bf109}, L140102 (2024).	
\bibitem{sfl9} Fang Qin, Ye Ma, Ruizhe Shen, and Ching Hua Lee, \textquotedblleft{Universal competitive spectral scaling from the critical non-Hermitian skin effect\textquotedblright}, Phys. Rev. B {\bf107}, 155430 (2023).
\bibitem{sfl10} S. M. Rafi-Ul-Islam, Zhuo Bin Siu, Haydar Sahin, Ching Hua Lee, and Mansoor B. A. Jalil, \textquotedblleft{Critical hybridization of skin modes in coupled non-Hermitian chains\textquotedblright}, Phys. Rev. Research {\bf4}, 013243 (2022).
\bibitem{sflcem} B Yilmaz, C Yuce, C Bulutay, \textquotedblleft{From scale-free to Anderson localization: a size-dependent transition\textquotedblright}, Phys. Rev. B {\bf110}, 214206 (2024).
\bibitem{sflwang} Cui-Xian Guo, Chun-Hui Liu, Xiao-Ming Zhao, Yanxia Liu, and Shu Chen, \textquotedblleft{Exact Solution of Non-Hermitian Systems with Generalized Boundary Conditions: Size-Dependent Boundary Effect and Fragility of the Skin Effect\textquotedblright}, Phys. Rev. Lett. {\bf127}, 116801 (2021).
\bibitem{scifun} S. Weidemann, M. Kremer, T. Helbig, T. Hofmann, A. Stegmaier, M. Greiter, R. Thomale, A. Szameit, \textquotedblleft{Topological funneling of light\textquotedblright}, Science {\bf6488}, 311 (2020).
\bibitem{yucefunnel} Z. Turker and C. Yuce, \textquotedblleft{The funneling eﬀect in a non-Hermitian Anderson Model\textquotedblright}, Phys. Scr. {\bf99}, 075028 (2024).
\bibitem{scifun2} Stefano Longhi, \textquotedblleft{Probing non-Hermitian skin effect and non-Bloch phase transitions\textquotedblright}, Phys. Rev. Research {\bf1}, 023013 (2019).	
\bibitem{scifun3} C. Yuce, \textquotedblleft{Spontaneous topological pumping in non-Hermitian systems \textquotedblright}, Phys. Rev. A {\bf99}, 032109 (2019).
\bibitem{in1} Li Wang, Qing Liu, and Yunbo Zhang, \textquotedblleft{Quantum dynamics on a lossy non-hermitian lattice\textquotedblright}, Chinese Phys. B {\bf 30}, 020506 (2021).
\bibitem{in2} Wen-Tan Xue, Yu-Min Hu, Fei Song, and Zhong Wang, \textquotedblleft{Non-Hermitian edge burst\textquotedblright}, Phys. Rev. Lett. {\bf 128}, 120401 (2022).
\bibitem{in3} C. Yuce and H. Ramezani, \textquotedblleft{Non-Hermitian edge burst without skin localization\textquotedblright}, Phys. Rev. B {\bf107}, L140302 (2023).	
\bibitem{in3cem} C. Yuce and H. Ramezani, \textquotedblleft{Strong edge burst with bipolar non-Hermitian skin effect\textquotedblright}, Phys. Rev. B {\bf109}, 214301 (2024).
\bibitem{in4} Lei Xiao, Wen-Tan Xue, Fei Song, Yu-Min Hu, Wei Yi, Zhong Wang, Peng Xue,  \textquotedblleft{Observation of non-Hermitian edge burst in quantum dynamics\textquotedblright}, 	Phys. Rev. Lett. {\bf133}, 070801 (2024).
\bibitem{offbea} Jiankun Zhu, Ya-Li Mao, Hu Chen, Kui-Xing Yang, Linhu Li, Bing Yang, Zheng-Da Li, and Jingyun Fan, \textquotedblleft{Observation of Non-Hermitian Edge Burst Effect in One-Dimensional Photonic Quantum Walk\textquotedblright}, Phys. Rev. Lett. {\bf132}, 203801 (2024).
\bibitem{offbea2}	Pengyu Wen, Jinghui Pi, and Gui-Lu Long, \textquotedblleft{Investigation of a non-Hermitian edge burst with time-dependent perturbation theory\textquotedblright}, Phys. Rev. A {\bf109}, 022236 (2024).
\bibitem{offbea3}	Shicheng Ma, Heng Lin, Jinghui Pi, \textquotedblleft{Imaginary gap-closed points and non-Hermitian dynamics in a class of dissipative systems\textquotedblright}, Phys. Rev. B {\bf109}, 214311 (2024).
\bibitem{bipolar2} K. Wang, A. Dutt, K. Y. Yang, C. C. Wojcik, J. Vuckovic, S. Fan, \textquotedblleft{Generating arbitrary topological windings of a non-Hermitian band\textquotedblright}, Science 371:1240 (2021).
    
\end{thebibliography}
\end{document}